\documentstyle[11pt,epsfig]{article}
\begin{document}
\title{\bf Cosmological Solutions of Chameleon Scalar Field Model }
\author{Raziyeh Zaregonbadi$^{1}$\thanks{email: r.zaregonbadi@malayeru.ac.ir}, Nasim Saba$^{2}$\thanks{email:
 n\_saba@sbu.ac.ir}$\, $ and Mehrdad Farhoudi$^{2}$\thanks{email: m-farhoudi@sbu.ac.ir} \\
 {\small $^{1}$Department of Physics, Faculty of Science, Malayer University, Malayer,
 Iran}\\
 {\small $^{2}$Department of Physics, Shahid Beheshti University, Evin, Tehran 19839,
 Iran}}
\date{\small  October 16, 2023 }
\maketitle
\begin{abstract}
\noindent
 We investigate cosmological solutions of the chameleon model
with a non-minimal coupling between the matter and the scalar
field through a conformal factor with gravitational strength. By
considering the spatially flat FLRW metric and the matter density
as a non-relativistic perfect fluid, we focus on the
matter-dominated phase and the late-time accelerated-phase of the
universe. In this regard, we manipulate and scrutinize the related
field equations for the density parameters of the matter and the
scalar fields with respect to the e-folding. Since the scalar
field fluctuations depend on the background and the field
equations become highly non-linear, we probe and derive the
governing equations in the context of various cases of the
relation between the kinetic and potential energies of the
chameleon scalar field, or indeed, for some specific cases of the
scalar field equation of state parameter. Thereupon, we
schematically plot those density parameters for two different
values of the chameleon non-minimal coupling parameter, and
discuss the results. In the both considered phases, we specify
that, when the kinetic energy of the chameleon scalar field is
much less than its potential energy (i.e., when the scalar field
equation of state parameter is $\simeq - 1 $), the behavior of the
chameleon model is similar to the $\Lambda CDM$ model. Such
compatibility suggests that the chameleon model is
phenomenologically viable and can be tested with the observational
data.
\end{abstract}


\section{Introduction}
\indent

It has been suggested that the universe starts from an extremely
rapid accelerated-phase after the Planck era called the
inflationary epoch that remedy important problems of the standard
cosmological model (the $\Lambda CDM$ model), see, e.g.,
Refs.~\cite{guth}-\cite{wein2}. The cosmic microwave background
observations are well consistent with the predictions of such an
inflation hypothesis~\cite{Plank}. In the literature, a lot of
work has been carried to investigate the inflationary scenario,
see, e.g., Refs.~\cite{nojiri2003}-\cite{Shiravand2022}. After the
inflationary epoch, observations predict that the universe has a
decelerating era during the early phase of its
evolution~\cite{Planck:2018}. In fact, the universe undergoes a
radiation dominated epoch followed by a matter dominated one.
Afterward, the present accelerated-phase of the universe starts
which is dubbed as a dark energy era. Indeed, observations -- such
as the type Ia supernova, the baryon acoustic oscillations, and
the gravitational waves -- confirm the late-time cosmic
acceleration~\cite{Plank,Planck:2018}-\cite{gw}. But the driver of
such acceleration has~not been identified by observations and has
no fully accepted theoretical model. However, since ordinary
matter cannot accelerate the universe, such dark energy (if it
exists) must be exotic matter for which many candidates have been
proposed and investigated. In most of these models, a scalar field
has been employed as dark sector with a dynamical equation of
state. On the other hand, modified gravitational theories (e.g.
Refs.~\cite{Farhoudi2006}-\cite{Duchaniya} and references therein)
have also been proposed mainly to describe the accelerated
evolution of the universe without any sector as dark energy. For
these two issues see, e.g., Refs.~\cite{carrolls}-\cite{Bajardi}
and references therein. Furthermore, other observations -- such as
the behavior of the galactic rotation curves and the mass
discrepancy in clusters of galaxies -- indicate the consideration
of an exotic matter called dark matter~\cite{Zwicky}-\cite{Binny}.
A lot of alternative efforts have also been performed on various
modifications to the Einstein field equations in order to deal
with the question of dark
matter~\cite{ra3,Borowiec,Milgrom}-\cite{zare3} and references
therein. Howsoever, the nature of both of these dark sectors
(which constitute about $95\% $ of the
universe~\cite{Plank,Planck:2018} and are well consistent with the
observational data~\cite{Ishak2019}) is one of the most important
issues in physics.

Although general relativity provides accurate predictions in
describing some cosmological phenomena~\cite{Will2006}, the
$\Lambda$CDM is the most acceptable model due to its high
conformity with the observational data of
cosmology~\cite{Ferreira2019}. However, this model still confronts
some challenges~\cite{weinb}-\cite{Bull} and other alternatives
have been proposed for it. The scalar-tensor theories of
gravitation, which extend general relativity by introducing a
scalar field, have become the most popular alternative to the
Einstein gravitational theory~\cite{wang}-\cite{bamba}.

In this work, we concentrate on the chameleon scalar field theory,
wherein unlike the common non-minimally coupled scalar field
theories, a chameleon field couples with matter sector rather than
the geometry. Such an interaction between the scalar field and
matter field causes the effective potential of the scalar field
depends on the matter density of the environment. Much work has
been done on this theory, see, e.g.,
Refs.~\cite{saba,screen2,sabazare,weltman1}-\cite{Paliathanasis2}
and references therein. This change of properties of the mass of
scalar field acts as a buffer with respect to observational
bounds. Accordingly, in this type of theory, a screening mechanism
is usually aimed at modifying general relativity on large scales.
Hence, the chameleon scalar field theory can be employed to
describe the accelerated expansion of the universe while the
effects of non-minimal coupling parameter are hidden in the small
scale gravitational experiments, see, e.g.,
Refs.~\cite{weltman1}-\cite{brax,khoury,brax2013}.

We have previously investigated the chameleon model during
inflation in Refs.~\cite{saba,saba1}, wherein we showed through
the proposed scenario that the effects of inflation and chameleon
can be described via a single scalar field during the inflation
and late-time. Also, we have probed the role of the chameleon
scalar field as dark energy in Ref.~\cite{sabazare}, where it was
claimed that such a model justifies dark energy with stronger
confirmation. Now, in this work, we intend to investigate the
cosmological solutions of the chameleon model in the hope that
these solutions can better justify the observational data. Of
course, since in the chameleon model, the properties of the scalar
fluctuations depend on the background, its field equations of
motion become highly non-linear. However, rather than solving the
resulted equations analytically, we restrict the solutions to
proceed. Indeed, for various cases of the relation between the
kinetic and potential energies of the chameleon scalar field (or
actually, for some specific cases of the scalar field equation of
state parameter), we solve the non-minimal coupled gravity
equations to understand the evolution of the scalar field.
Meanwhile, some attempts have been done to find analytic solutions
for the chameleon model, wherein these analytic solutions of the
chameleon model exist only for highly symmetric source shapes such
as spheres, plates, and ellipses~\cite{weltman1,a3,a2}). If the
shape of the matter source is irregular, it will be more difficult
to find analytical solutions to the equations of motion.
Nevertheless, in Ref.~\cite{a4}, a software package called SELCIE
has been introduced that provides tools for constructing an
arbitrary system of mass distributions and then computing the
corresponding solution to the chameleon scalar field equation. In
addition, through the dynamical systems
technique~\cite{Wainwright,Coley}, one can also study the full
non-linearity of these cosmological models. In this technique,
usually normalized dimensionless new variables are introduced,
then by finding the fixed/critical points of the system and their
stability, the evolution of the system can be pictured
qualitatively near these points. The dynamical systems technique
has also been performed for the chameleon scalar field, see, e.g.,
Refs.~\cite{Farajollahi2011,RoyBanerjee,Roy2015,Paliathanasis}.

The work is organized as follows. In the next two sections, while
benefiting from our previous works~\cite{saba,sabazare,saba1}, we
first introduce the chameleon model and then, by considering the
spatially flat Friedmann-Lema\^{\i}tre-Robertson-Walker (FLRW)
metric, we obtain the corresponding field equations of motion.
Thereupon, in Sec.~III, we manipulate and scrutinize the related
field equations with respect to the e-folding for better and more
use in the later sections. In Secs.~IV and V, the cosmological
solutions of the chameleon model are respectively investigated for
the matter-dominated phase and the late-time accelerated-phase of
the universe for various cases of the relation between the kinetic
and potential energies of the chameleon scalar field. Indeed, we
probe the results of the chameleon model for some specific cases
of the scalar field equation of state parameter. Meanwhile, in
Sec.~V, we take a brief look at the chameleon scalar field
profile. At last, we conclude the work in Sec.~VI with the summary
of the results.

\section{Chameleon Model with Scalar Field}\label{Chameleon}
\indent

We consider an action of the chameleon model with a scalar field
in four dimensions when there are a minimal coupling between the
Einstein gravity and that scalar field, and a non-minimal coupling
between matter species with it as
\begin{eqnarray}\label{action}
S\!\!\!&=\!\!\!&\!\!\int{\rm
d}^{4}x\sqrt{-g}\left(\frac{M^{2}_{\rm
Pl}R}{2}\right)\!-\!\!\!\int{\rm
d}^{4}x\sqrt{-g}\left[\frac{1}{2}\partial_{\mu}\phi
\partial^{\mu}\phi+V(\phi)\right]\nonumber
\\
 &&+\sum_{i}\int{\rm d}^{4}x\sqrt{-{\tilde g}_{(i)}}{L_{(i)}^{\rm (m)}}\left({\psi}_{(i)}^{\rm (m)},{\tilde
 g}_{(i)\mu\nu}\right).
\end{eqnarray}
In this action, $\phi$ is a scalar field, $V(\phi)$ is a
self-interacting potential, $\psi_{(i)}$s are various matter
fields, ${L^{\rm (m)}_{(i)}}$s are Lagrangians of matter fields,
$\tilde g_{(i)\mu\nu}$s are matter field metrics that are
conformally related to the Einstein frame metric as
\begin{equation}
{\tilde g}_{(i)\mu \nu}={e}^{ {2\frac {{\beta}
_{(i)}\,\phi}{M_{\rm Pl}}}}{g}_{\mu \nu}.
\end{equation}
Here, $\beta_{(i)}$s are dimensionless constants that represent
different non-minimal coupling parameter between the scalar field
and each matter species, however in this work, we only consider a
single matter component. Also, the lower case Greek indices run
from zero to three, $g$ is the determinant of the metric, $R$ is
the Ricci scalar, and the reduced Planck mass is $M_{\rm
Pl}\equiv(8\pi G)^{-1/2}\approx 10^{27}eV$ in the natural units of
$\hbar=1=c$. Typically in the literature, the common power-law
chameleon potential
\begin{equation}\label{usual-potential}
V\left( \phi  \right) = \frac{{{M^{4 + n}}}}{{{\phi ^n}}}
\end{equation}
is usually used, where $n$ is a positive or negative integer
constant\footnote{Consistent values of $\beta$ with the allowed
integers $n$ have been mentioned in, e.g.,
Refs.~\cite{Burrage2017,Burrage2016}.}\
 and $M$ is some positive constant mass scale.

The variation of action (\ref{action}) with respect to the scalar
field gives the field equation
\begin{equation}\label{bax}
\Box\phi=\frac{{dV\left( \phi  \right)}}{{d\phi
}}-\frac{{\beta}}{M_{\rm Pl}}e^{4\frac{{\beta}\phi}{{M}_{\rm
Pl}}}{\tilde g}^{\mu\nu}{\tilde T}_{\mu\nu}^{(m)},
\end{equation}
where the box symbol $\Box\equiv\nabla^{\alpha}\nabla_{\alpha}$
corresponds to the metric $g_{\mu\nu}$, and $\tilde
T_{\mu\nu}^{(m)}$ is the energy-momentum tensor of matter (which
is conserved in the Jordan frame) defined as
\begin{equation}\label{tj}
\tilde T_{\mu\nu}^{(m)}=-\frac{2}{\sqrt{-\tilde
{g}}}\frac{\left(\delta{\sqrt{-\tilde
{g}}}{L^{(m)}}\right)}{\delta {\tilde g}^{\mu\nu}}.
\end{equation}
Also, the variation with respect to the
metric $g_{\mu \nu}$ associated to the Einstein frame gives
\begin{equation}\label{G}
{G_{\mu \nu }} = \frac{1}{M_{\rm Pl}^2}\left(T_{\mu \nu
}^{(\phi)}+T_{\mu \nu }^{(m)}\right)= \frac{1}{M_{\rm
Pl}^2}\left(T_{\mu \nu }^{(\phi)}+ {e}^{ {2\frac
{{\beta}\phi}{M_{\rm Pl}}}}\tilde T_{\mu\nu}^{(m)}\right),
\end{equation}
where $ T_{\mu \nu }^{(\phi)} $ is the energy-momentum tensor of
the scalar field as
\begin{equation}\label{phiii}
T_{\mu \nu }^{(\phi)}  =  - \frac{1}{2}{g_{\mu \nu }}{\partial
^\alpha } \phi {\partial _\alpha }\phi  - {g_{\mu \nu }}V\left(
\phi  \right) + {\partial _\mu }\phi {\partial _\nu }\phi.
\end{equation}
In addition, we assume the matter field as a perfect fluid in the
Jordan frame with the linear barotropic equation of state $\tilde
p^{(m)}=w \tilde \rho^{(m)}$. Hence, one obtains the trace of the
energy-momentum tensor (\ref{tj}), with the signature $+2$, as
\begin{equation}
\tilde{T}^{(m)}={\tilde g}^{\mu\nu}\tilde
T_{\mu\nu}^{(m)}=-(1-3w)\tilde \rho^{(m)},
\end{equation}
where the relation between the matter density in the Einstein
frame with the Jordan frame is
\begin{equation}\label{rhoo}
\rho^{(m)}= e^{4\frac{{\beta}\phi}{M_{\rm Pl}}}\tilde \rho^{(m)}.
\end{equation}

The matter density $\tilde\rho^{(m)}$ is conserved in the Jordan
frame, i.e.
\begin{equation}\label{conservRhTilde}
\dot{\tilde\rho}^{(m)}+3\tilde
H\left(1+w\right)\tilde\rho^{(m)}=0,
\end{equation}
however, it is~not conserved in the Einstein frame. Nevertheless,
the mathematical quantity
\begin{equation}\label{rh}
\rho(t)\equiv e^{3(1+w)\frac{{\beta}\phi}{M_{\rm Pl}}}\tilde
\rho^{(m)},
\end{equation}
independent of the scalar field and with the same equation of
state parameter, is a conserved quantity in this frame, i.e.
\begin{equation}\label{conservRh}
\dot{\rho}+3H\left(1+w\right)\rho=0.
\end{equation}
Of course, this $\rho(t)$ is~not a physical matter density.
Furthermore, substituting definition~(\ref{rh}) into
Eq.~(\ref{bax}) yields the dynamic of the scalar field governed by
an effective potential, i.e.
\begin{equation}\label{field}
\Box\phi=\frac{{d{V_{\rm eff}}\left(\phi\right)}}{{d\phi }},
\end{equation}
where
\begin{equation}\label{eff}
{V}_{\rm eff}(\phi)\equiv V(\phi)+\rho
e^{(1-3w)\frac{\beta\phi}{M_{\rm Pl}}}=V(\phi)+\rho^{(m)},
\end{equation}
which depends explicitly on the matter density. Additionally, the
mass of the chameleon field, which is sufficiently large to evade
local constraints~\cite{weltman1,Khoury2010}, is
\begin{equation}
m_{\rm eff,\min }^2 = {V''_{\rm eff}}\left( {{\phi _{\min }}}
\right) = V''\left( {{\phi _{\min }}} \right) + \frac{{{\beta ^2}
}}{{M_{\rm Pl}^2}} {\left( {1 - 3w} \right)^2}\rho
{e^{\frac{{\beta {\phi _{\min }}}}{{{M_{\rm Pl}}}}}}.
\end{equation}

In the following, we derive the relevant field equations in order
to investigate the cosmology of such a chameleon model.

\section{Scrutinizing Field Equations}\label{Cosmological}
\indent

In this section, we intend to manipulate and scrutinize the
cosmological equations of the chameleon model while considering
the spatially flat FLRW metric in the Einstein frame, namely
\begin{equation}\label{metric}
ds^{2}=-dt^{2}+a^{2}(t)\left(dx^{2}+dy^{2}+dz^{2}\right),
\end{equation}
where $a(t)$ is the scale factor as a function of the cosmic time
$t$. In this respect, first, the field equation (\ref{field})
gives the wave equation
\begin{equation}\label{phi}
\ddot{\phi}+3H\dot\phi+\frac{{d{V_{\rm eff}}\left(\phi\right)}}{{d\phi }} = 0,
\end{equation}
where we have also considered the scalar field just as a function
of the cosmic time, dot denotes the derivative with respect to
this time, and $H(t)\equiv\dot{a}/a$ is the Hubble parameter.
Furthermore by metric (\ref{metric}), Eq. (\ref{G}) gives the
corresponding Friedmann and Raychaudhuri equations as
\begin{equation}\label{friedmann}
{H^2} = \frac{1}{3{M_{\rm Pl}^2}}\left( {\frac{1}{2}{{\dot \phi
}^2} + V\left( \phi  \right) + \rho {e^{\left( {1 - 3w}
\right)\frac{{\beta \phi }}{{{M_{\rm Pl}}}}}}} \right)
\end{equation}
and
\begin{equation}\label{fried22}
-M^{2}_{\rm Pl}\left(\frac{2\ddot{a}}{a}+H^{2}\right)=
\frac{1}{2}\dot{\phi}^{2}-V(\phi)+w\rho
e^{\left(1-3w\right)\frac{\beta\phi}{M_{\rm Pl}}}.
\end{equation}
Also, relation (\ref{phiii}) gives the energy and the pressure
densities as
\begin{equation}\label{rho p phi}
{\rho ^{\left( \phi  \right)}} = \frac{1}{2}{{\dot \phi }^2} +
V\left( \phi  \right)\qquad {\rm and}\qquad {p^{\left( \phi
\right)}} = \frac{1}{2}{{\dot \phi }^2} - V\left( \phi  \right).
\end{equation}

Accordingly, Eqs.~(\ref{friedmann}) and (\ref{fried22}) can be
rewritten as
\begin{equation}\label{friedman}
{H^2} = \frac{1}{{3M_{\rm Pl}^2}}\left( {{\rho ^{\left(\phi
\right)}} + {\rho ^{\left( m \right)}}} \right) =
\frac{1}{{3M_{\rm Pl}^2}} {\rho ^{\left( {\rm tot} \right)}},
\end{equation}
\begin{equation}\label{fried2}
\frac{{\ddot a}}{a} =  - \frac{1}{{6M_{\rm Pl}^2}}\left[ {{\rho
^{\left( \phi  \right)}} + {\rho ^{\left( m \right)}} + 3\left(
{{p^{\left( \phi  \right)}} + {p^{\left( m \right)}}} \right)}
\right] =  - \frac{1}{{6M_{\rm Pl}^2}}\left( {{\rho ^{\left( {\rm
tot} \right)}} + 3{p^{\left( {\rm tot} \right)}}} \right),
\end{equation}
where $\rho^{(\rm tot)} \equiv \rho^{(\phi)}+\rho^{(m)}$ and
$p^{(\rm tot)} \equiv p^{(\phi)}+p^{(m)}$ are the total energy and
pressure densities. The field equations (\ref{friedman}) and
(\ref{fried2}) indicate that $\rho^{(\rm tot)}$ is conserved in
the Einstein frame, i.e.
\begin{equation}\label{rhotot}
{\dot{\rho}}^{(\rm tot)}+3H\left(\rho^{(\rm tot)}+p^{(\rm
tot)}\right)=0.
\end{equation}
However, in general, the scalar and matter fields are~not
separately conserved in this frame, and an interacting term stands
among those, i.e.
\begin{equation}\label{rhom}
{\dot{\rho}}^{(m)}+3H\left(\rho^{(m)}+p^{(m)}\right)=\frac{\beta\left(1-3w\right)}{M_{\rm
Pl}}\rho^{(m)}\dot{\phi}
\end{equation}
and
\begin{equation}\label{rhophi}
{\dot{\rho}}^{(\phi)}+3H\left(\rho^{(\phi)}+p^{(\phi)}\right)=-\frac{\beta\left(1-3w\right)}{M_{\rm
Pl}}\rho^{(m)}\dot{\phi},
\end{equation}
wherein the interacting term depends on the scalar field and the
background matter density of the environment. However, for an
ultra-relativistic matter, with $w=1/3$, such an interacting term
is identically zero. That is, for the radiation dominated epoch of
the universe, the scalar and matter fields are separately
conserved.

In the following analysis, we consider the matter density as a
non-relativistic perfect fluid\rlap,\footnote{We mean all types of
matter, i.e. baryonic and non-baryonic matter (including dark
matter). }\
 i.e. dust matter with $w=0$.
Accordingly, Eqs.~(\ref{rhom}) and (\ref{rhophi}) read
\begin{equation}\label{rhh1}
{\dot{\rho}}^{(m)}+3H\rho^{(m)}=\frac{\beta}{M_{\rm Pl}}\rho^{(m)}\dot{\phi}
\end{equation}
and
\begin{equation}\label{rhh2}
{\dot{\rho}}^{(\phi)}+ 3H{\rho ^{(\phi)} }\left( {1 + {w^{(\phi)}
}} \right) =- \frac{\beta}{M_{\rm Pl}}\rho^{(m)}\dot{\phi},
\end{equation}
where ${w^{(\phi)}}=p ^{(\phi)}/\rho ^{(\phi)}$. Moreover, we
present the equations via the dimensionless density parameters
defined as
\begin{equation}\label{om0}
{\Omega^{(m)}} \equiv \frac{\rho^{(m)}}{{\rho _0^{\rm (crit)}}}
\qquad {\rm and} \qquad {\Omega^{(\phi)} } \equiv
\frac{{{\rho^{(\phi)} }}}{{\rho _0^{\rm (crit)}}},
\end{equation}
where $ \rho _0^{\rm (crit)} \equiv 3H_0^2M_{\rm Pl}^2 $ is the
critical density of the universe at the present-time. Hence, Eq.
(\ref{friedman}) can be rewritten as
\begin{equation}
{H^2} = H_0^2\left[ { {\Omega^{(\phi)} }+{\Omega^{(m)}}} \right],
\end{equation}
which in turn imposes constraint
\begin{equation}\label{ConstrainCondition}
{\Omega_0^{(\phi)} }+{\Omega_0^{(m)}}=1
\end{equation}
on the initial values at the present-time.

On the other hand, by employing the e-folding variable
\begin{equation}\label{Nwrta}
N = \ln \left[ {\frac{{a(t)}}{{a({t_0})}}} \right],
\end{equation}
we have
\begin{equation}
\frac{d}{{dt}} = H\frac{d}{{dN}}.
\end{equation}
Also from relations~(\ref{rho p phi}) and (\ref{om0}),  we obtain
\begin{equation}\label{31}
\dot{\phi}=\pm\sqrt {{\rho ^{\left( \phi  \right)}} + {p^{\left(
\phi \right)}}}  = \pm{H_0}{M_{\rm Pl}}\sqrt {3{\Omega^{(\phi)}
}\left( {1 + {w^{(\phi)} }} \right)}.
\end{equation}
It has been shown~\cite{Khoury2010} that the monotonic increase of
the matter coupling factor $ {e^{\frac{{\beta \phi }}{{{M_{\rm
Pl}}}}}} $, when corresponds to $ \dot \phi  > 0 $, leads to a
minimum for ${V_{\rm eff}}\left( \phi \right) $. Moreover, It has
been stated~\cite{Roshan} that $ \dot \phi  > 0 $ is satisfactory
from the astrophysical point of view~\cite{Ferreira1997}. Hence,
we confine our investigation\footnote{We neglect the oscillatory
case of the chameleon scalar field.}\
 to the positive sign in relation (\ref{31}) to proceed, and
the generalization to the negative sign can be performed with
similar considerations. Thus, Eqs. (\ref{rhh1}) and (\ref{rhh2})
read
\begin{equation}\label{om1}
{\Omega '}^{(m)}+ 3{\Omega^{(m)}} =
\frac{{\beta {H_0}}}{H} {\Omega^{(m)}}
\sqrt {3{\Omega^{(\phi)} }\left( 1 + w^{(\phi) } \right)}
\end{equation}
and
\begin{equation}\label{om2}
{{\Omega '}^{(\phi)} } + 3{\Omega^{(\phi)} }\left( {1 + {w^{(\phi)} }} \right) = -\frac{{\beta {H_0}}}{H} {\Omega^{(m)}}
\sqrt {3{\Omega^{(\phi)} }\left( {1 + {w^{(\phi)} }} \right)},
\end{equation}
where the prime denotes the derivative with respect to the
e-folding.

In addition, in the case of $w=0$, the defined total equation of
state parameter for this model is
\begin{equation}\label{WT}
{w^{\left( {\rm tot} \right)}} \equiv
\frac{{{p^{\left( {\rm tot} \right)}}}}
{{{\rho ^{\left( {\rm tot} \right)}}}} =
\frac{{{p^{\left( \phi  \right)}}}}{{{\rho ^{\left( m \right)}}
+ {\rho ^{\left( \phi  \right)}}}}.
\end{equation}
Also, using the deceleration parameter, i.e. $q \equiv  - {\ddot
a}a/{\dot a}^2 = - {\ddot a}/(aH^2)$, while substituting
Eqs.~(\ref{friedman}) and (\ref{fried2}), with employing the
definition of the total equation of state parameter, into it,
leads to
\begin{equation}\label{q2}
q = \frac{{1 + 3{w^{\left( {\rm tot} \right)}}}}{2}.
\end{equation}
Obviously if $ q<0 $ (which is equal to $ {w^{\left( {\rm tot}
\right)}}<-1/3 $), it will describe an accelerated-phase in
evolution of the universe.

In the continuation of the work, we intend to investigate the
cosmological solutions of the chameleon model for two important
phases of the cosmos.

\section{Matter-Dominated Phase}
\indent

Under assumption ${\rho^{\left(m \right)}} \gg {\rho^{\left(\phi
\right)}} $, which corresponds to $ {\Omega^{(m)}} \gg {\Omega
^{(\phi)}}$, the total equation of state parameter~(\ref{WT})
yields
\begin{equation}\label{wm}
{w^{\left( {\rm tot} \right)}} \simeq \frac{{{p ^{\left( \phi
\right)}}}}{{{\rho ^{\left( m \right)}}}}.
\end{equation}
Now, if the chameleon potential $V\left( \phi \right)$ being
positive, then relations (\ref{rho p phi}) will indicate that $
{p^{\left( \phi \right)}} < {\rho ^{\left( \phi \right)}} $, and
in turn ${\rho ^{\left( m \right)}} \gg {p^{\left( \phi  \right)}}
$, hence relation~(\ref{wm}) gives
\begin{equation}\label{zero-total}
{w^{\left( {\rm tot} \right)}} \simeq 0,
\end{equation}
for the matter density as a non-relativistic perfect fluid.
Furthermore, in this situation, the deceleration
parameter~(\ref{q2}) is
\begin{equation}\label{negative-era}
q \simeq \frac{1}{2},
\end{equation}
i.e. the expansion of the universe is a decelerated evolution.

On the other hand, in the matter-dominated phase, Eq.
(\ref{friedman}) reduces to
\begin{equation}\label{M}
{H^2} \simeq H_0^2{\Omega^{(m)} },
\end{equation}
and hence, Eqs.~(\ref{om1}) and (\ref{om2}) read
\begin{equation}\label{om11}
{\Omega '}^{(m)}+ 3{\Omega^{(m)}} \simeq \beta \sqrt
{3\Omega^{(m)} \,\, {\Omega^{(\phi)} } {\left( {1 + {w^{(\phi)} }}
\right)}}
\end{equation}
and
\begin{equation}\label{om22}
{{\Omega '}^{(\phi)} } + 3{\Omega^{(\phi)} }\left( {1 +
{w^{(\phi)} }} \right) \simeq -\beta \sqrt {3\Omega^{(m)} \,\,
{\Omega^{(\phi)}}{\left( {1 + {w^{(\phi)} }}\right)}}.
\end{equation}
In the above, one obviously has $\rho^{(m)}> 0$ and $H>0$, wherein
we have assumed that\footnote{Note that, by considering relations
(\ref{rho p phi}) plus the case $p^{(\phi)}=w^{(\phi)}
\rho^{(\phi)}$, it leads that $\rho^{(\phi)}$ cannot be zero.}\
 $\rho^{(\phi)}> 0$, i.e. ${\dot \phi}^2/2> \mid
V(\phi)\mid$ for negative potentials. Also, we have assumed that
\begin{equation}
{w^{(\phi)} }\geq -1,
\end{equation}
which, with relations (\ref{rho p phi}) and the assumption
$\rho^{(\phi)}> 0$, it is already satisfied.

At this stage, it is more instructive to define two new parameters
$ y \equiv \sqrt {{\Omega ^{\left( m \right)}}}$ and $ x \equiv
\sqrt {{\Omega ^{\left( \phi \right)}}}$. Thus, we can rewrite
Eqs.~(\ref{om11}) and (\ref{om22}) as the following coupled first
derivative equations for $x$ and $y$, namely
\begin{equation}\label{y}
2y' + 3y \simeq \beta x\sqrt {3{\left( {1 + {w^{(\phi)}
}}\right)}}
\end{equation}
and
\begin{equation}\label{x}
2x' + 3x{\left( {1 + {w^{(\phi)} }}\right)}\simeq - \beta y\sqrt
{3{\left( {1 + {w^{(\phi)} }}\right)}}.
\end{equation}
However, rather than solving these coupled equations analytically,
we restrict the solutions to proceed. Indeed, in the following
subsections, we investigate the solutions of these coupled
differential equations for various cases of the relation between
the kinetic and potential energies of the chameleon scalar field.
Actually, we probe the results of the chameleon model for four
specific cases of the scalar field equation of state parameter.

\subsection{Case ${\dot \phi ^2}/2 \gg \mid V\left( \phi  \right)\mid $}
\indent

In this case, one obviously has $ {w^{\left( \phi  \right)}}
\simeq + 1 $, hence Eqs.~(\ref{y}) and (\ref{x}) read
\begin{equation}\label{y2}
2y' + 3y \simeq \sqrt{6}\,\beta x,
\end{equation}
\begin{equation}\label{x2}
2x' + 6x \simeq -\sqrt{6}\,\beta y.
\end{equation}
Now, by applying the upper bound value of the $\beta=3.7\times
{10^2} $ (which this value of the non-minimal coupling parameter
in the chameleon model has been shown~\cite{Jaffe} to be
consistent with the experimental constrain), the solutions of
these equations are
\begin{eqnarray}\label{46}
y\left( N \right) =  - \frac{{{e^{ -
\frac{9}{4}N}}}}{{1480}}\Bigg[
 \!\!\!\!\!&\Big( &\!\!\!\!\! {C_1} \sqrt 6  -
{C_2}\sqrt {2190394}  \Big)\sin \left( {\frac{{\sqrt {3285591}
}}{4}N} \right)\cr
 \!\!\!\!\!&+&\!\!\!\!\! \left( {{C_1}\sqrt {2190394}  + {C_2}\sqrt 6
} \right)\cos \left( {\frac{{\sqrt {3285591} }}{4}N} \right)
\Bigg]
\end{eqnarray}
and
\begin{equation}\label{47}
x\left( N \right) = {e^{ - \frac{9}{4}N}}\left[ {{C_1}\sin \left(
{\frac{{\sqrt {3285591} }}{4}N} \right) + {C_2}\cos \left(
{\frac{{\sqrt {3285591} }}{4}N} \right)} \right],
\end{equation}
where $C_1$ and $C_2=x(0)$ are constants of integrations. Using
solutions (\ref{46}) and (\ref{47}), we have plotted $ {\Omega
^{\left( m \right)}} = {y^2} $ and ${\Omega ^{\left( \phi
\right)}} = {x^2} $ in Figure~1, wherein (without loss of
generality) the present values of the parameters have been used to
specify the constants\rlap.\footnote{Note that, since the
chameleon scalar field can cause the late-time accelerated
expansion of the universe (see, e.g.,
Refs.~\cite{sabazare,Banerjee2010}), we have used the data of dark
energy for it. Besides, although we consider the matter-dominated
phase, we have used the initial values at the present-time because
the path of the functions eventually passes through this point.
Moreover, the current value of the matter density is still a
suitable value even at almost higher redshift from the
present-time~\cite{capozziello}. }\
 This figure indicates
that the behavior of the matter density and the chameleon scalar
field density are both damped oscillations, i.e. the value of both
densities decreases over time. Note that, the negative values of
the e-folding in this figure (and also in the subsequent figures)
are because we have assumed the value of the scale factor to be
equal to one at the present-time, i.e. ${a({t_0})}=1 $. We should
also remind that the zero values of ${\Omega^{\left( m \right)}}$
and ${\Omega^{\left( \phi \right)}} $ must be disregarded in all
the figures.

On the other hand, by considering the non-minimal coupling
parameter as $ \beta=1 $, Eqs.~(\ref{y2}) and (\ref{x2}) lead to
solutions
\begin{equation}\label{48}
y\left( N \right) =  - \frac{{e^{ - \frac{9}{4}N}}}{4}\left[
{\left( {{C_3} \sqrt 6  - {C_4}\sqrt {10} } \right)\sin \left(
{\frac{{\sqrt {15} }}{4}N} \right) + \left( {{C_3}\sqrt {10}  +
{C_4}\sqrt 6 } \right)\cos \left( {\frac{{\sqrt {15} }}{4}N}
\right)} \right]
\end{equation}
and
\begin{equation}\label{49}
x\left( N \right) = {e^{ - \frac{9}{4}N}}\left[ {{C_3}\sin \left(
{\frac{{\sqrt {15} }}{4}N} \right) + {C_4}\cos \left(
{\frac{{\sqrt {15} }}{4}N} \right)} \right],
\end{equation}
where $C_3$ and $C_4=x(0)$ are constants of integrations. Also,
using solutions (\ref{48}) and (\ref{49}), we have plotted $
{\Omega^{\left( m \right)}}$ and ${\Omega ^{\left( \phi \right)}}
$ in Figure~2. This figure indicates that at the beginning of this
phase of the universe the matter density dominates, and then at
the end of its dominance era, the chameleon scalar field density
is dominant.
%%%%%%%%%%%%%%%%%%%%%%%%%%%%%%%%%%%%%%%%%%%%%%%%%%%%%%%%%%%%%%%%%%
\begin{figure}[h]
\centerline{\includegraphics[scale=0.7]{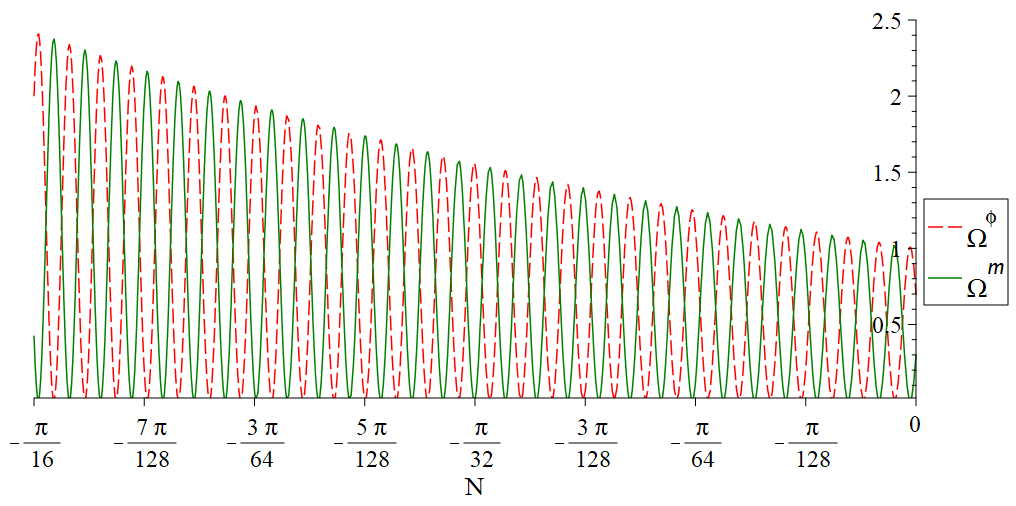}}
\caption{\label{1figure} For solutions (\ref{46}) and (\ref{47})
with $ \beta=3.7\times {10^2} $, the figure schematically (i.e.,
scale-free) shows the parameters $ {\Omega ^{\left( m \right)}} $
and ${\Omega ^{\left( \phi \right)}}$ versus $N$ as solid and
dashed lines, respectively. Besides, the present values of the
parameters, i.e. $ \Omega^{(m)} _{0} \simeq {0.3} $ and $
{\Omega^{(\phi)} _{0}} \simeq {0.7} $ (the current observational
data for all matter is up to $32\% $ and for dark energy roughly
$68\% $ of the universe~\cite{Planck:2018}), have been used as the
initial conditions, and ${a({t_0})}=1 $.}
\end{figure}
%%%%%%%%%%%%%%%%%%%%%%%%%%%%%%%%%%%%%%%%%%%%%%%%%%%%%%%%%%%%%%%%%%
\begin{figure}[h]
\centerline{\includegraphics[scale=0.7]{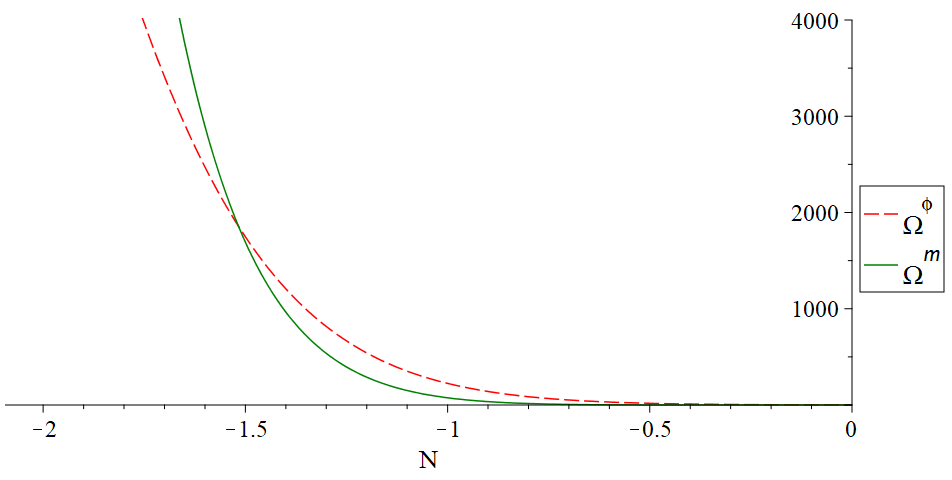}}
\caption{\label{2figure} For solutions (\ref{48}) and (\ref{49})
with $ \beta=1 $, the figure schematically (i.e., scale-free)
shows the parameters $ {\Omega ^{\left( m \right)}} $ and ${\Omega
^{\left( \phi  \right)}}$ versus $N$ as solid and dashed lines,
respectively. Besides, the present values of the parameters, i.e.
$ \Omega^{(m)} _{0} \simeq {0.3} $ and $ {\Omega^{(\phi)} _{0}}
\simeq {0.7} $, have been used as the initial conditions, and
${a({t_0})}=1 $.}
\end{figure}
%%%%%%%%%%%%%%%%%%%%%%%%%%%%%%%%%%%%%%%%%%%%%%%%%%%%%%%%%%%%%%

\subsection{Case ${\dot \phi ^2}/2 = V\left( \phi  \right) $}
\indent

In this case, the chameleon potential must be positive, and
accordingly relations (\ref{zero-total}) and (\ref{negative-era})
will hold. In addition, one gets $ {w^{\left( \phi \right)}} =0 $,
hence Eqs.~(\ref{y}) and (\ref{x}) read
\begin{equation}\label{y3}
2y' + 3y \simeq \sqrt{3}\,\beta x,
\end{equation}
\begin{equation}\label{x3}
{}\quad 2x' + 3x \simeq -\sqrt{3}\,\beta y.
\end{equation}
The solutions to these equations with $ \beta=3.7\times {10^2} $
are
\begin{equation}\label{52}
y\left( N \right) =  - {e^{ - \frac{3}{2}N}}\left[ {{C_5}\cos
\left( {185\sqrt 3 N} \right) - {C_6}\sin \left( {185\sqrt 3 N}
\right)} \right]
\end{equation}
and
\begin{equation}\label{53}
x\left( N \right) = {e^{ - \frac{3}{2}N}}\left[ {{C_5}\sin \left(
{185\sqrt 3 N} \right) + {C_6}\cos \left( {185\sqrt 3 N} \right)}
\right],
\end{equation}
where $C_5=-y(0)$ and $C_6=x(0)$ are constants of integrations.
Using solutions (\ref{52}) and (\ref{53}), we have plotted $
{\Omega^{\left( m \right)}} $ and ${\Omega^{\left( \phi \right)}}
$ in Figure~3.
%%%%%%%%%%%%%%%%%%%%%%%%%%%%%%%%%%%%%%%%%%%%%%%%%%%%%%%%%%%%%%%%%%
\begin{figure}[h]
\centerline{\includegraphics[scale=0.7]{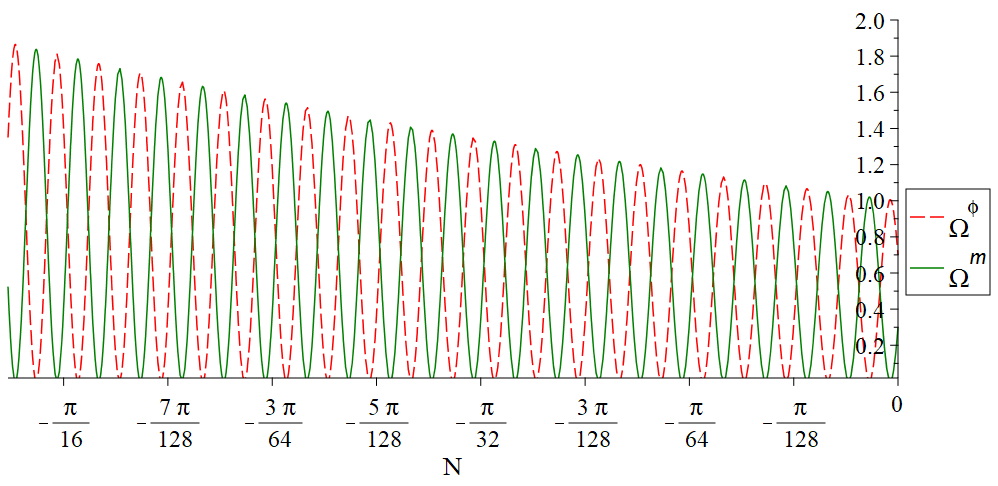}}
\caption{\label{3figure} For solutions (\ref{52}) and (\ref{53})
with $ \beta=3.7\times {10^2} $, the figure schematically (i.e.,
scale-free) shows the parameters $ {\Omega ^{\left( m \right)}} $
and ${\Omega ^{\left( \phi \right)}}$ versus $N$ as solid and
dashed lines, respectively. Besides, the present values of the
parameters, i.e. $ \Omega^{(m)} _{0} \simeq {0.3} $ and $
{\Omega^{(\phi)} _{0}} \simeq {0.7} $, have been used as the
initial conditions, and  ${a({t_0})}=1 $.}
\end{figure}
%%%%%%%%%%%%%%%%%%%%%%%%%%%%%%%%%%%%%%%%%%%%%%%%%%%%%%%%%%%%%%
This figure indicates that the behavior of the matter density and
the chameleon scalar field density are both also damped
oscillations and their values decrease over time.

However, using the non-minimal coupling parameter $ \beta=1 $,
Eqs.~(\ref{y3}) and (\ref{x3}) lead to
\begin{equation}\label{54}
y\left( N \right) =  - {e^{ - \frac{3}{2}N}}\left[ {{C_7}\cos
\left( {\frac{{\sqrt 3 }}{2}N} \right) - {C_8}\sin \left(
{\frac{{\sqrt 3 }}{2}N} \right)} \right]
\end{equation}
and
\begin{equation}\label{55}
x\left( N \right) = {e^{ - \frac{3}{2}N}}\left[ {{C_7}\sin \left(
{\frac{{\sqrt 3 }}{2}N} \right) + {C_8}\cos \left( {\frac{{\sqrt 3
}}{2}N} \right)} \right],
\end{equation}
where $C_7=-y(0)$ and $C_8=x(0)$ are constants of integrations.
Again, we have plotted $ {\Omega ^{\left( m \right)}} $ and
${\Omega ^{\left( \phi \right)}} $ from solutions (\ref{54}) and
(\ref{55}) in Figure~4.
%%%%%%%%%%%%%%%%%%%%%%%%%%%%%%%%%%%%%%%%%%%%%%%%%%%%%%%%%%%%%%%%%%
\begin{figure}[h]
\centerline{\includegraphics[scale=0.8]{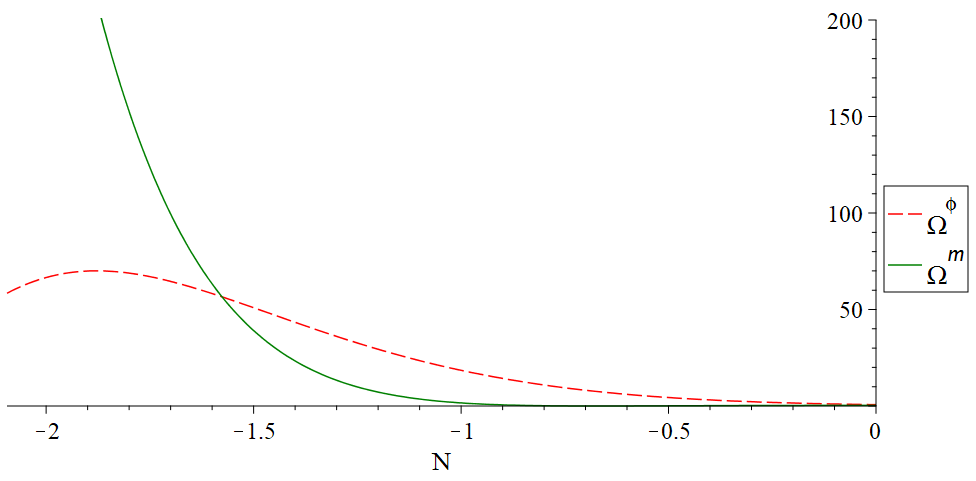}}
\caption{\label{4figure} For solutions (\ref{54}) and (\ref{55})
with $ \beta=1 $, the figure schematically (i.e., scale-free)
shows the parameters $ {\Omega ^{\left( m \right)}} $ and ${\Omega
^{\left( \phi  \right)}}$ versus $N$ as solid and dashed lines,
respectively. Besides, the present values of the parameters, i.e.
$ \Omega^{(m)} _{0} \simeq {0.3} $ and $ {\Omega^{(\phi)} _{0}}
\simeq {0.7} $, have been used as the initial conditions, and
${a({t_0})}=1 $.}
\end{figure}
%%%%%%%%%%%%%%%%%%%%%%%%%%%%%%%%%%%%%%%%%%%%%%%%%%%%%%%%%%%%%%
This figure also indicates that at the beginning of this phase the
matter density dominates, and then at the end of its dominance
era, the chameleon scalar field density is dominant.

\subsection{Case ${\dot \phi ^2}/2 = V\left( \phi  \right)/2 $}
\indent

In this case, the chameleon potential must again be positive, and
hence relations (\ref{zero-total}) and (\ref{negative-era}) will
hold. Although, one gets $ {w^{\left( \phi  \right)}} =-1/3 $,
that is why we have highlighted this case, which also brings
almost different results. Accordingly, Eqs.~(\ref{y}) and
(\ref{x}) read
\begin{equation}\label{y4}
2y' + 3y \simeq \sqrt{2}\,\beta x,
\end{equation}
\begin{equation}\label{x4}
{}\quad 2x' + 2x \simeq -\sqrt{2}\,\beta y.
\end{equation}
The solutions to these equations with $ \beta=3.7\times {10^2} $
are
\begin{equation}\label{58}
y\left( N \right)\! = \frac{{{e^{ -
\frac{5}{4}N}}}}{{1480}}\!\left[ {\left( {7{C_{10}} \sqrt {44702}
-\! {C_9}\sqrt 2 } \right)\sin\! \left( {\frac{{7\sqrt {22351}
}}{4}N} \right)\! - \!\left( {7{C_9}\sqrt {44702}  -\!
{C_{10}}\sqrt 2 } \right)\cos\! \left( {\frac{{7\sqrt {22351}
}}{4}N} \right)} \right]
\end{equation}
and
\begin{equation}\label{59}
x\left( N \right) = {e^{ - \frac{5}{4}N}}\left[ {{C_9}\sin \left(
{\frac{{7 \sqrt {22351} }}{4}N} \right) + {C_{10}}\cos \left(
{\frac{{7\sqrt {22351} }}{4}N} \right)} \right],
\end{equation}
where $C_9$ and $C_{10}=x(0)$ are constants of integrations. This
time, using solutions (\ref{58}) and (\ref{59}), we have plotted
${\Omega ^{\left( m \right)}}$ and ${\Omega ^{\left( \phi
\right)}} $ in Figure~5.
%%%%%%%%%%%%%%%%%%%%%%%%%%%%%%%%%%%%%%%%%%%%%%%%%%%%%%%%%%%%%%%%%%
\begin{figure}[h]
\centerline{\includegraphics[scale=0.7]{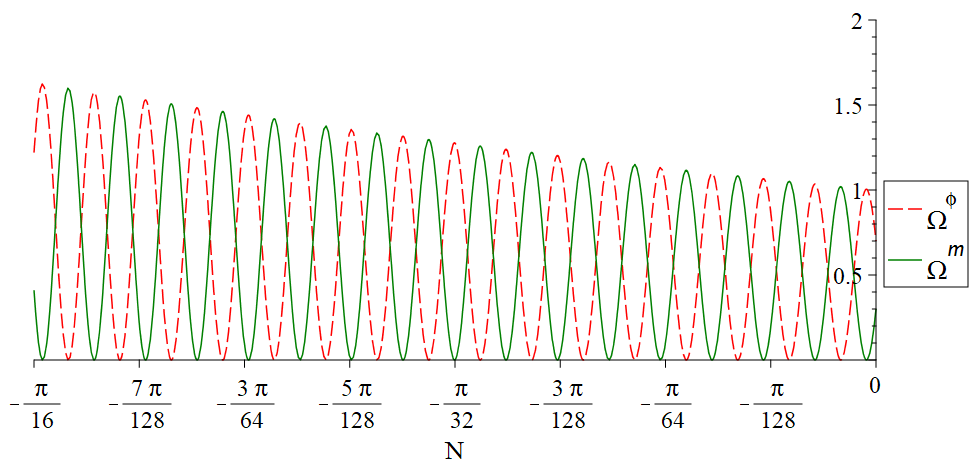}}
\caption{\label{5figure} For solutions (\ref{58}) and (\ref{59})
with $ \beta=3.7\times {10^2} $, the figure schematically (i.e.,
scale-free) shows the parameters $ {\Omega ^{\left( m \right)}} $
and ${\Omega ^{\left( \phi \right)}}$ versus $N$ as solid and
dashed lines, respectively. Besides, the present values of the
parameters, i.e. $ \Omega^{(m)} _{0} \simeq {0.3} $ and $
{\Omega^{(\phi)} _{0}} \simeq {0.7} $, have been used as the
initial conditions, and  ${a({t_0})}=1 $.}
\end{figure}
%%%%%%%%%%%%%%%%%%%%%%%%%%%%%%%%%%%%%%%%%%%%%%%%%%%%%%%%%%%%%%
This figure indicates that the behavior of the matter density and
the chameleon scalar field density are both also damped
oscillations and their values decrease over time.

On the other hand, by considering the non-minimal coupling
parameter as $ \beta=1 $, Eqs.~(\ref{y4}) and (\ref{x4}) lead to
solutions
\begin{equation}\label{60}
y\left( N \right) =  - \frac{{{e^{ - \frac{5}{4}N}}}}{4}\left[
{\left( {{C_{11}}\sqrt 2 + {C_{12}}\sqrt {14} } \right)\sin \left(
{\frac{{\sqrt 7 }}{4}N} \right) + \left( {{C_{12}} \sqrt 2  -
{C_{11}}\sqrt {14} } \right)\cos \left( {\frac{{\sqrt 7 }}{4}N}
\right)} \right]
\end{equation}
and
\begin{equation}\label{61}
x\left( N \right) = {e^{ - \frac{5}{4}N}}\left[ {{C_{11}}\sin
\left( {\frac{{\sqrt 7 }}{4}N} \right) + {C_{12}}\cos \left(
{\frac{{\sqrt 7 }}{4}N} \right)} \right],
\end{equation}
where $C_{11}$ and $C_{12}=x(0)$ are constants of integrations.
Again using solutions (\ref{60}) and (\ref{61}), we have plotted $
{\Omega^{\left( m \right)}}$ and ${\Omega ^{\left(\phi \right)}}$
in Figure~6.
%%%%%%%%%%%%%%%%%%%%%%%%%%%%%%%%%%%%%%%%%%%%%%%%%%%%%%%%%%%%%%%%%%
\begin{figure}[h]
\centerline{\includegraphics[scale=0.8]{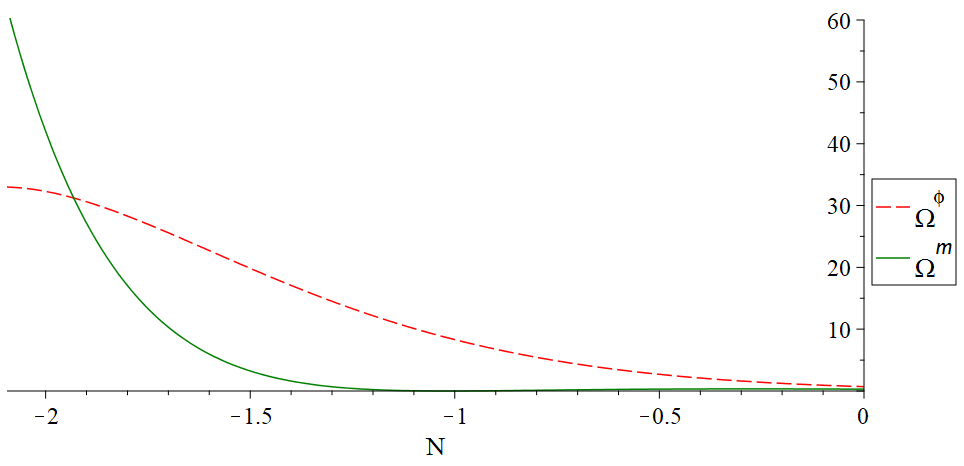}}
\caption{\label{6figure} For solutions (\ref{60}) and (\ref{61})
with $ \beta=1 $, the figure schematically (i.e., scale-free)
shows the parameters $ {\Omega ^{\left( m \right)}} $ and ${\Omega
^{\left( \phi  \right)}}$ versus $N$ as solid and dashed lines,
respectively. Besides, the present values of the parameters, i.e.
$ \Omega^{(m)} _{0} \simeq {0.3} $ and $ {\Omega^{(\phi)} _{0}}
\simeq {0.7} $, have been used as the initial conditions, and
${a({t_0})}=1 $.}
\end{figure}
%%%%%%%%%%%%%%%%%%%%%%%%%%%%%%%%%%%%%%%%%%%%%%%%%%%%%%%%%%%%%%
This figure also shows that at the beginning of this phase, the
matter density dominates as assumed, but at the end of its
dominance era, the chameleon scalar field density is dominant.

\subsection{Case ${\dot \phi ^2}/2 \ll V\left( \phi  \right)$}
\indent

Once again, in this case, the chameleon potential must be
positive\rlap,\footnote{Note that, as we have assumed
$\rho^{(\phi)}>0$, we do~not consider the case ${\dot \phi ^2}/2
\ll \mid V\left( \phi  \right)\mid$ for negative potentials.}\
 and accordingly relations
(\ref{zero-total}) and (\ref{negative-era}) will hold. Moreover,
we obviously have $ {w^{\left( \phi \right)}} \simeq -1 $, hence
Eqs.~(\ref{y}) and (\ref{x}) read
\begin{equation}\label{y5}
2y' + 3y \simeq 0,
\end{equation}
\begin{equation}\label{x5}
2x'\simeq 0.
\end{equation}
These equations do~not depend on the $ \beta $ parameter, and we
have the solutions
\begin{equation}\label{64}
y\left( N \right) = {y\left( 0 \right)}{e^{ - \frac{3}{2}N}}
\end{equation}
and
\begin{equation}\label{65}
x\left( N \right) = {x\left( 0 \right)},
\end{equation}
where $y\left( 0 \right) = \sqrt {\Omega _0^{\left( m \right)}}  $
and $x\left( 0 \right)=\sqrt {\Omega _{0 }^{\left( \phi \right)}}
$ are the initial conditions at the present-time\rlap.\footnote{
Considering relation (\ref{Nwrta}), $N=0$ is obviously related to
$t_0$ and in turn to $z=0$. }\
 Once again, using solutions (\ref{64}) and (\ref{65}), we have
plotted $ {\Omega ^{\left( m \right)}}$ and ${\Omega^{\left( \phi
\right)}}$ in Figure~7.
%%%%%%%%%%%%%%%%%%%%%%%%%%%%%%%%%%%%%%%%%%%%%%%%%%%%%%%%%%%%%%%%%%
\begin{figure}[h]
\centerline{\includegraphics[scale=0.8]{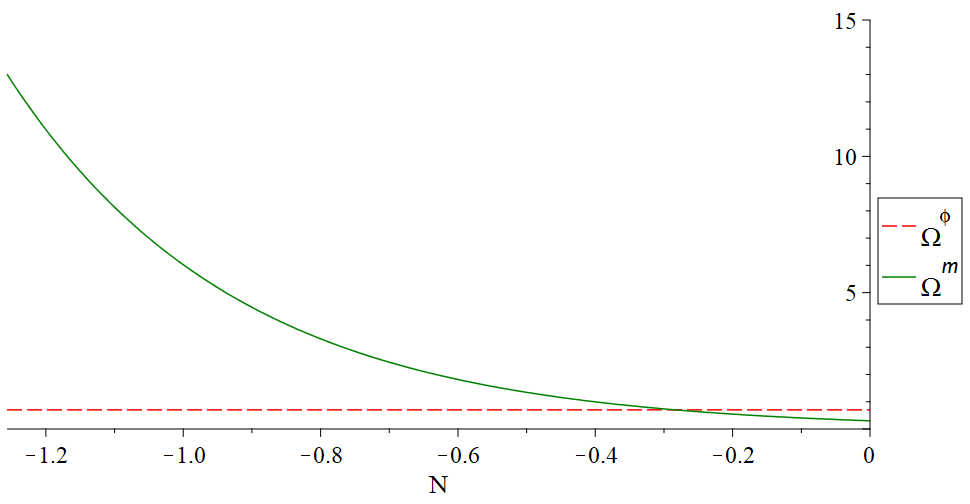}}
\caption{\label{7figure} For solutions (\ref{64}) and (\ref{65}),
the figure schematically (i.e., scale-free) shows the parameters $
{\Omega ^{\left( m \right)}} $ and ${\Omega ^{\left( \phi
\right)}}$ versus $N$ as solid and dashed lines, respectively.
Besides, the present values of the parameters, i.e. $ \Omega^{(m)}
_{0} \simeq {0.3} $ and $ {\Omega^{(\phi)} _{0}} \simeq {0.7} $,
have been used as the initial conditions, and  ${a({t_0})}=1 $.}
\end{figure}
%%%%%%%%%%%%%%%%%%%%%%%%%%%%%%%%%%%%%%%%%%%%%%%%%%%%%%%%%%%%%%
This figure also indicates that at the beginning of this phase,
the matter density dominates as assumed, but at the end of its
dominance era, the constant density of the chameleon scalar field
is dominant.

\section{Late-Time Accelerated-Phase}
\indent

In this section, we investigate the evaluation of the late-time
accelerated-phase of the universe via the chameleon scalar field.
In this regard, in this accelerated-phase of the universe, by
reducing the dust matter density over time, we can assume $ {\rho
^{\left(\phi \right)}} \gg {\rho ^{\left( m \right)}} $. Hence,
the equation of state parameter~(\ref{WT}) reduces to
\begin{equation}\label{wphii}
{w^{\left( {\rm tot} \right)}} \simeq \frac{{{p^{\left( \phi
\right)}}}}{{{\rho ^{\left( \phi  \right)}}}} = {w^{\left( \phi
\right)}},
\end{equation}
which gives the deceleration parameter~(\ref{q2}) as $q \simeq
\left({1 + 3{w^{\left( \phi  \right)}}}\right)/2$. Then, in order
to achieve $ q \le 0 $ in this phase of the universe, one must use
$ {w^{\left( \phi \right)}}\le -1/3 $, which leads to the
constraint ${\dot \phi ^2} \le V\left( \phi  \right) $ with
positive potentials in the chameleon model.

On the other hand, in the accelerated-phase of the universe, Eq.
(\ref{friedman}) yields
\begin{equation}\label{H}
{H^2} \simeq H_0^2{\Omega^{(\phi)} }.
\end{equation}
Substituting Eq. (\ref{H}) into Eqs. (\ref{om1}) and (\ref{om2})
gives
\begin{equation}\label{m}
{{\Omega '}^{(m)}} + 3{\Omega^{(m)}} \simeq \beta {\Omega^{(m)}}
\sqrt {3\left( {1 + {w^{(\phi)} }} \right)}
\end{equation}
and
\begin{equation}\label{omegap}
{{\Omega '}^{(\phi)} } + 3{\Omega^{(\phi)} }\left( {1 +
{w^{(\phi)} }} \right) \simeq -\beta {\Omega^{(m)}} \sqrt {3\left(
{1 + {w^{(\phi)} }} \right)}.
\end{equation}

Once again, in the following subsections, we probe these solutions
for various cases of the relation between the kinetic and
potential energies of the scalar field that satisfy the ${\dot
\phi ^2}/2 \le V\left( \phi \right)/2 $ constraint with positive
potentials in the chameleon model. Meanwhile, while using the
common power-law chameleon potential (\ref{usual-potential}), we
take a brief look at the chameleon scalar field profile in this
phase.

\subsection{Case ${\dot \phi ^2}/2 = V\left( \phi  \right)/2 $}
\indent

In this case, we have $ {w^{\left( \phi  \right)}} =-1/3 $, and in
turn $ q=0 $, which is the transition point between the
decelerated-epoch and the accelerated-phase in the evaluation of
the universe. Hence, Eqs.~(\ref{m}) and (\ref{omegap}) with $
\beta=3.7\times {10^2} $ read
\begin{equation}
\frac{{{{\Omega '}^{\left( m \right)}}}} {{{\Omega ^{\left( m
\right)}}}}\simeq 3.7\times {10^2} \sqrt 2 - 3
\end{equation}
and
\begin{equation}
\frac{{{{\Omega '}^{\left( \phi  \right)}}}}{{{\Omega ^{\left(
\phi  \right)}}}} \simeq  - 3.7 \times {10^2}\sqrt 2 \left(
{\frac{{{\Omega ^{\left( m \right)}}}}{{{\Omega ^{\left( \phi
\right)}}}}} \right) - 2.
\end{equation}
The solutions to these equations are
\begin{equation}\label{m5.1.1}
{{\Omega^{(m)}}}(N) = {\Omega^{(m)} _{ 0 }} {e^{\left({3.7\times
{10^2}\sqrt{2}}-3 \right)N}}
\end{equation}
and
\begin{equation}\label{phi5.1.1}
{\Omega ^{\left( \phi  \right)}}(N) = \left\{ {\Omega _0^{\left(
\phi \right)} + \frac{{3.7 \times {{10}^2}\sqrt 2\, \Omega
_0^{\left( m \right)}}}{{273799}}\left( {1 + 3.7 \times
{{10}^2}\sqrt 2 } \right)\left[ {1 - {e^{\left( {3.7 \times
{{10}^2}\sqrt 2  - 1} \right)N}}} \right]} \right\}{e^{ - 2N}}.
\end{equation}
We have plotted $ {\Omega ^{\left( m \right)}} $ and ${\Omega
^{\left( \phi \right)}}$ in Figure~8.
%%%%%%%%%%%%%%%%%%%%%%%%%%%%%%%%%%%%%%%%%%%%%%%%%%%%%%%%%%%%%%%%%%
\begin{figure}[h]
\centerline{\includegraphics[scale=0.8]{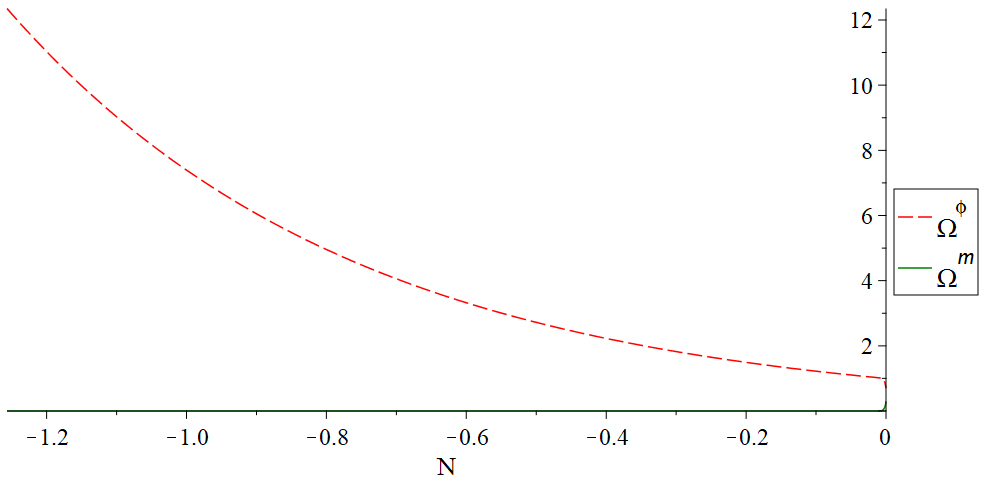}}
\caption{\label{8figure} For solutions (\ref{m5.1.1}) and
(\ref{phi5.1.1}) with $\beta=3.7\times {10^2} $, the figure
schematically (i.e., scale-free) shows the parameters $ {\Omega
^{\left( m \right)}} $ and ${\Omega ^{\left( \phi \right)}}$
versus $N$ as solid and dashed lines, respectively. Besides, the
present values of the parameters, i.e. $ \Omega^{(m)} _{0} \simeq
{0.3} $ and $ {\Omega^{(\phi)} _{0}} \simeq {0.7} $, have been
used as the initial conditions, and  ${a({t_0})}=1 $.}
\end{figure}
%%%%%%%%%%%%%%%%%%%%%%%%%%%%%%%%%%%%%%%%%%%%%%%%%%%%%%%%%%%%%%
This figure shows that, in this phase of the universe, although
the matter density increases slightly, its amount is very
negligible all along the path, and the chameleon scalar field
density, while decreasing, is dominant at all times.

On the other hand, for $ \beta=1 $, Eq.~(\ref{m}) gives
\begin{equation}\label{m5.1}
\frac{{{{\Omega '}^{\left( m \right)}}}} {{{\Omega ^{\left( m
\right)}}}} \simeq \sqrt 2 - 3
\end{equation}
and, while using the assumption $ {\Omega ^{\left( m \right)}} \ll
{\Omega^{\left(\phi  \right)}} $ in the late-time
accelerated-phase of the universe, Eq.~(\ref{omegap}) yields
\begin{equation}\label{omegapp}
\frac{{{{\Omega '}^{\left(\phi  \right)}}}}{{{\Omega ^{\left( \phi
\right)}}}} + 2 \simeq 0.
\end{equation}
The solutions to Eqs. (\ref{m5.1}) and (\ref{omegapp}) are
\begin{equation}\label{rhoo4}
{{\Omega^{(m)}}}(N) = {\Omega^{(m)} _{ 0 }} {e^{\left({\sqrt{2}}-3
\right)N}}
\end{equation}
and
\begin{equation}\label{omegap4}
{\Omega^{(\phi)} }(N) = \Omega_{0}^{\left(\phi  \right)} {e^{ - 2
N}}.
\end{equation}
We have plotted $ {\Omega ^{\left( m \right)}} $ and ${\Omega
^{\left( \phi \right)}}$ in Figure~9.
%%%%%%%%%%%%%%%%%%%%%%%%%%%%%%%%%%%%%%%%%%%%%%%%%%%%%%%%%%%%%%%%%%
\begin{figure}[h]
\centerline{\includegraphics[scale=0.8]{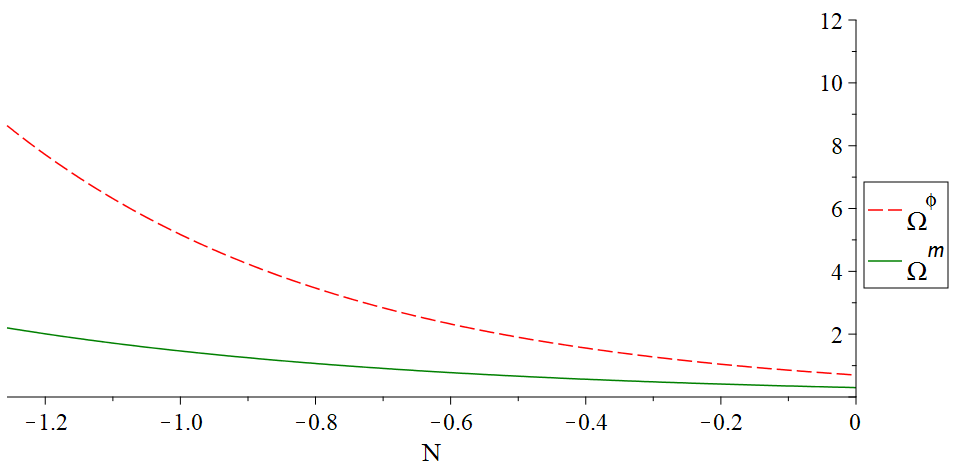}}
\caption{\label{9figure} For solutions (\ref{rhoo4}) and
(\ref{omegap4}) with $\beta= 1 $, the figure schematically (i.e.,
scale-free) shows the parameters $ {\Omega ^{\left( m \right)}} $
and ${\Omega ^{\left( \phi \right)}}$ versus $N$ as solid and
dashed lines, respectively. Besides, the present values of the
parameters, i.e. $ \Omega^{(m)} _{0} \simeq {0.3} $ and $
{\Omega^{(\phi)} _{0}} \simeq {0.7} $, have been used as the
initial conditions, and ${a({t_0})}=1 $.}
\end{figure}
%%%%%%%%%%%%%%%%%%%%%%%%%%%%%%%%%%%%%%%%%%%%%%%%%%%%%%%%%%%%%%
This figure indicates that the non-relativistic matter density and
the chameleon scalar field density both decrease with increasing
the scale factor. Although the decrease in the matter density
during this epoch of the evolution of the universe is less than
that of the chameleon, the chameleon scalar field density
dominates at all times.

Continuing the analysis, let us take a brief look at the chameleon
scalar field profile while using the common power-law chameleon
potential (\ref{usual-potential}). In the case of this subsection,
such a potential leads to
\begin{equation}
\dot \phi  =  \pm \sqrt {\frac{{{M^{4 + n}}}}{{{\phi ^n}}}}.
\end{equation}
Therefore, the chameleon scalar field is
\begin{equation}
{\phi^{\frac{n}{2} + 1}}(t) =  \pm \left( {\frac{n}{2} + 1}
\right) \sqrt {{M^{4 + n}}}\, t + {A(t=0)},
\end{equation}
where $ {A(t=0)} $ is an integration constant. Furthermore, the
constraint $ V\left( \phi  \right) > 0 $ indicates that, for the
case $ \phi >0 $, the parameter $n$ can be both odd and even, but
for case $ \phi < 0 $, the parameter $n$ can only be even.

\subsection{Case ${\dot \phi ^2}/2 \ll V\left( \phi  \right) $}
\indent

In this case, we have $ {w^{\left( \phi  \right)}} \simeq -1 $,
and in turn $ q<0 $, hence Eqs.~(\ref{m}) and (\ref{omegap}) read
\begin{equation}
{\Omega '^{\left( m \right)}} + 3{\Omega ^{\left( m \right)}}
\simeq 0
\end{equation}
and
\begin{equation}
{\Omega '^{\left( \phi  \right)}} \simeq 0.
\end{equation}
These equations do~not depend on the $ \beta $ parameter, and we
have the solutions
\begin{equation}\label{rhoo5}
{{\Omega^{(m)}}}(N) = {\Omega^{(m)} _{0}} {e^{ -3 N}}
\end{equation}
and
\begin{equation}\label{omegap5}
{\Omega^{(\phi)} }(N) = \Omega _{0}^{\left(\phi  \right)}.
\end{equation}
We have plotted solutions (\ref{rhoo5}) and (\ref{omegap5}) in
Figure~10.
%%%%%%%%%%%%%%%%%%%%%%%%%%%%%%%%%%%%%%%%%%%%%%%%%%%%%%%%%%%%%%%%%%
\begin{figure}[h]
\centerline{\includegraphics[scale=0.8]{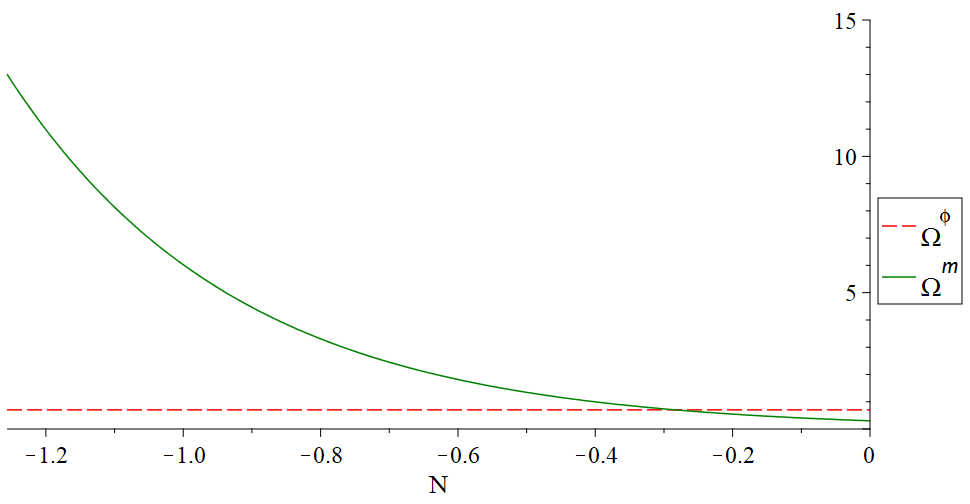}}
\caption{\label{10figure} For solutions (\ref{rhoo5}) and
(\ref{omegap5}), the figure schematically (i.e., scale-free) shows
the parameters $ {\Omega ^{\left( m \right)}} $ and ${\Omega
^{\left( \phi \right)}}$ versus $N$ as solid and dashed lines,
respectively. Besides, the present values of the parameters, i.e.
$ \Omega^{(m)} _{0} \simeq {0.3} $ and $ {\Omega^{(\phi)} _{0}}
\simeq {0.7} $, have been used as the initial conditions, and
${a({t_0})}=1 $.}
\end{figure}
%%%%%%%%%%%%%%%%%%%%%%%%%%%%%%%%%%%%%%%%%%%%%%%%%%%%%%%%%%%%%%
This figure illustrates that first the matter density dominates,
then with the increase of the scale factor, the chameleon scalar
field density, while always remaining constant, dominates the
late-time accelerated-phase of the universe.

Also, in the case of subsection $5.2$, the common power-law
chameleon potential (\ref{usual-potential}) leads to
\begin{equation}
\dot \phi  \ll \pm \sqrt {\frac{{{2M^{4 + n}}}}{{{\phi ^n}}}}.
\end{equation}
Hence, the chameleon scalar field is
\begin{equation}
{\phi ^{\frac{n}{2} + 1}}(t) \ll  \pm \left( {\frac{n}{2} + 1}
\right) \sqrt {{2M^{4 + n}}}\, t + {B(t=0)},
\end{equation}
where $ {B(t=0)} $ is an integration constant.

Now, to make our investigations more instructive, let us compare
our results with the corresponding one in the $\Lambda CDM$ model.
For the $\Lambda CDM$ model, we have
\begin{equation}
S = \int {{d^4}x\sqrt { - g} } \left[ {\frac{{M_{\rm
Pl}^2}}{2}\left( {R - 2\Lambda } \right) + {L^{\left( m
\right)}}\left( {{\psi ^{\left( m \right)}},{g_{\mu \nu }}}
\right)} \right],
\end{equation}
with the field equations
\begin{equation}
{G_{\mu \nu }} = \frac{1}{{M_{\rm Pl}^2}}\left( {T_{\mu \nu
}^{\left( \Lambda  \right)} + T_{\mu \nu }^{\left( m \right)}}
\right),
\end{equation}
where the energy-momentum tensor of the cosmological constant is
defined as
\begin{equation}
T_{\mu \nu }^{\left( \Lambda  \right)}\equiv  - M_{\rm
Pl}^2\Lambda {g_{\mu \nu }}.
\end{equation}
This tensor describes a vacuum state with a constant energy
density $ {\rho ^{\left( \Lambda  \right)}} $ and a constant
isotropic pressure density $ {p^{\left( \Lambda  \right)}} $ as
\begin{equation}
{\rho ^{\left( \Lambda  \right)}} = M_{\rm Pl}^2\Lambda = -{p
^{\left( \Lambda  \right)}}.
\end{equation}
Assuming the FLRW metric (\ref{metric}) with the matter density as
a dust matter perfect fluid, one obtains the Friedmann equations
for the $\Lambda CDM$ model as
\begin{equation}
{H^2} = \frac{\Lambda }{3} + \frac{{{\rho ^{\left( m
\right)}}}}{{3M_{\rm Pl}^2}},
\end{equation}
\begin{equation}
\frac{{\ddot a}}{a} = \frac{\Lambda }{3} - \frac{{{\rho ^{\left( m
\right)}}}}{{6M_{\rm Pl}^2}}.
\end{equation}
Furthermore, the energy-momentum conservation in this model leads
to
\begin{equation}\label{Rho}
{{\dot \rho }^{\left( m \right)}} + 3H\rho ^{\left( m \right)}= 0,
\end{equation}
\begin{equation}\label{Rho L}
{{\dot \rho }^{\left( \Lambda  \right)}} = 0.
\end{equation}
Using the dimensionless density parameters, these are
\begin{equation}
{{\Omega '}^{\left( m \right)}} + 3{\Omega ^{\left( m \right)}} =
0,
\end{equation}
\begin{equation}
{{\Omega '}^{\left( \Lambda  \right)}} = 0
\end{equation}
that lead to solution (\ref{rhoo5}) and solution like
(\ref{omegap5}), respectively.

Therefore, in the case $ {\dot \phi ^2}/2 \ll V\left( \phi \right)
$ with positive potentials in the chameleon model, i.e.
subsections $4.4$ and $5.2$, the behavior obtained for the
presented chameleon model is similar to the $\Lambda CDM$ model.

\section{Conclusions}\label{Sec6}
\indent

By considering the spatially flat FLRW line element as the
background geometry, we have investigated the cosmological
solutions of the chameleon model. In this model, the scalar field
non-minimally couples with the matter field, and its interaction
with the ambient matter goes through a conformal factor that leads
to a dependence of the chameleon mass on the matter density.
Accordingly, the equations of motion of the chameleon scalar field
become highly non-linear, hence rather than solving the resulted
equations analytically, we have restricted the solutions to
proceed.

After deriving the corresponding Friedmann and Raychaudhuri
equations, we have manipulated and scrutinized the related field
equations for the dimensionless density parameters of the matter
field and the scalar field with respect to the e-folding while
treating the matter density as a non-relativistic perfect fluid
and all variables simply as a function of the cosmic time. Then,
we have focused and investigated the cosmological solutions of the
chameleon model in the matter-dominated phase and the late-time
accelerated-phase of the universe in the context of various cases
of the relation between the kinetic and potential energies of the
chameleon scalar field, or indeed, for some specific cases of the
scalar field equation of state parameter. Thereupon, we have
schematically plotted those density parameters versus the
e-folding for two different values on the chameleon non-minimal
coupling parameter. Meanwhile, we have shown that the assumption
of the matter density much more than the chameleon density
corresponds to the matter-dominated phase with a decelerated
evolution for positive chameleon potential energies. However
conversely, the assumption of the chameleon density much more than
the matter density can correspond to an accelerated-phase when the
scalar field equation of state parameter is less than or equal to
$ -1/3$ (in this case, the equality corresponds to the
deceleration parameter being zero, i.e. the transition point).
This situation leads to the case that the kinetic energy of the
chameleon scalar field must be less than or equal to half of its
potential energy for positive chameleon potentials.

More clearly, for the described chameleon model in the
matter-dominated phase, with the non-minimal coupling parameter
$\beta=1 $, we have indicated that, when the kinetic energy of the
chameleon scalar field is much more than the absolute value of,
equal to, and equal to half of its potential
energy\footnote{Obviously, in the last two cases, the potential
energy must be positive.}\
 (which
respectively correspond to the scalar field equation of state
parameter of approximately $ + 1 $, $0 $, and $-1/3$), the matter
density dominates at the beginning, but then, at the end of its
dominance era, the chameleon scalar field density is dominant.
Conversely, if the kinetic energy of the chameleon scalar field is
much less than its potential energy for positive chameleon
potential energies (i.e., when the scalar field equation of state
parameter is approximately $ - 1 $), the final result will still
be the same as above, but for any non-minimal coupling parameter.
That is, the result of this case does~not depend on the
non-minimal coupling parameter. Also, in this phase with the
strong non-minimal coupling parameter $\beta=3.7\times {10^2} $,
the behavior of the matter density and the chameleon scalar field
density, in the first three cases mentioned above, are both damped
oscillations, i.e. the value of both densities decreases with
increasing scale factor over time.

Furthermore, in the late-time accelerated-phase, first, the
plausible assumption of the chameleon density much more than the
matter density leads to the total equation of state parameter
being almost identical to the scalar field equation of state
parameter, the one described at the end of the second paragraph.
Then, in the case that the kinetic energy of the chameleon scalar
field is equal to half of its potential energy, with the strong
non-minimal coupling parameter $\beta=3.7\times {10^2} $, we have
indicated that, although the matter density increases slightly,
its amount is very negligible all along the path and the chameleon
scalar field density, while decreasing, is dominant at all times.
In this case with the non-minimal coupling parameter $\beta=1 $,
the matter density and the chameleon scalar field density both
decrease, and although the decrease of the matter density is less
than that of the chameleon, the chameleon scalar field density
dominates at all times. On the other hand, in the case that the
kinetic energy of the chameleon scalar field is much less than its
potential energy (i.e., when the scalar field equation of state
parameter is approximately $ - 1 $), the behavior of the matter
density and the chameleon scalar field density do~not depend on
the non-minimal coupling parameter. In this case, we have shown
that first the matter density dominates, then with the increase of
the scale factor, the chameleon scalar field density, while always
remaining constant, dominates the late-time phase. Also, we have
taken a brief look at the chameleon scalar field profile while
using the common power-law chameleon potential.

Finally, in both the matter-dominated phase and the late-time
accelerated-phase, we have specified that, when the kinetic energy
of the chameleon scalar field is much less than its potential
energy (i.e., when the scalar field equation of state parameter is
approximately $ - 1 $), the behavior obtained for the presented
chameleon model is similar to the $\Lambda CDM$ model. Such
compatibility suggests that the chameleon model is
phenomenologically viable and can be tested with the observational
data.

%%%%%%%%%%%%%%%%%%%%%%%%%%%%%%%%%%%%%%%%%%%%%%%%%%%%%%%%%%%%%%%%%%%%%%%

\end{document}